\documentclass{article}
\usepackage[T1]{fontenc} % add special characters (e.g., umlaute)
\usepackage[utf8]{inputenc} % set utf-8 as default input encoding
\usepackage{ismir,amsmath,cite,url}
\usepackage{graphicx}
\usepackage{color}
\usepackage{booktabs}
\usepackage{multirow}
\usepackage{subfigure}
\usepackage{tabularx}

\title{Automatic Composition of Guitar Tabs by Transformers and Groove Modeling }

\multauthor{Yu-Hua Chen$^{1,2,3}$,~ Yu-Hsiang Huang$^1$,~ Wen-Yi Hsiao$^1$,~ and Yi-Hsuan Yang$^{1,2}$} {  
  $^1$ Taiwan AI Labs, Taiwan,~
  $^2$ Academia Sinica, Taiwan,~
  $^3$ National Taiwan University, Taiwan\\
{\tt\small r08946011@ntu.edu.tw, \{yshuang,wayne391,yhyang\}@ailabs.tw }
}

\begin{document}

\maketitle

\begin{abstract}
Deep learning algorithms are increasingly developed for learning to compose music in the form of MIDI files.  However, whether such algorithms work well for composing guitar tabs, which are quite different from MIDIs, remain relatively unexplored. To address this, we build a model for composing fingerstyle guitar tabs with Transformer-XL, a neural sequence model architecture. With this model, we investigate the following research questions. First, whether the neural net generates note sequences with meaningful  note-string combinations, which is important for the guitar but not other instruments such as the piano. Second, whether it generates compositions with coherent rhythmic groove, crucial for fingerstyle guitar music. And, finally, how pleasant the composed music is in comparison to real, human-made compositions. Our work provides preliminary empirical evidence of the promise of deep learning for tab composition, and suggests areas for future study.
% Thanks to the cumulative efforts in the community, recent years have witnessed great progress in automatic composition of music in symbolic formats such as the piano rolls and MIDIs.  Automatic composition of the guitar tabs, however, remains relatively unexplored to date.  To address this discrepancy, this paper proposes a methodology to convert a guitar tab to a series of discrete event tokens, facilitating the use of modern neural sequence models such as the Transformer-XL to model and to generate tabs as event sequences.  Our event representation entails consideration of the fingering (i.e., string-fret combinations) and the right-hand playing techniques associated with the musical notes, in an endeavor to make the generated tab playable. Moreover, we consider different approaches to model the rhythmic grooving of guitar for use in a neural sequence model. We train a tab composition model using a collection of fingerstyle guitar tabs, and validate the effectiveness of our design with both objective metrics and subjective user study.
\end{abstract}

\section{Introduction } 
\label{sec:introduction}

Thanks to the cumulative efforts in the community, in recent years we have seen great progress in using deep learning models for automatic music composition \cite{briot19book}. An important body of research has been invested on creating piano compositions, or more generally keyboard style music. For instance, the ``Music Transformer''  presented by Huang \emph{et al.}  \cite{huang2018music} employs  172 hours of piano performances to learn to compose classical piano music. Another group of researchers extends that model to generate pop piano compositions from 48 hours of human-performed piano covers \cite{huang2020pop}. They both use a MIDI-derived representation of music and describe music as a sequence of event tokens such as \texttt{NOTE-ON} and \texttt{NOTE-VELOCITY}.
%\footnote{MIDI, which stands for Musical Instrument Digital Interface, is a communication protocol designed for digital musical instruments.} %\cite{oore2018time}.} 
While 
%the staff notation assumed in 
the MIDI format works the best for representing keyboard instruments and less for other instruments %such as guitar and strings 
(for reasons described below),  Donahue \emph{et al.} \cite{donahue2019lakhnes} and Payne \cite{musenet} show respectively that it is possible for machines to learn from a set of MIDI files to compose multi-instrument music.
%a neural sequence model that is self-claimed to be able to generate compelling minute-long classical piano compositions.

\begin{figure}[t]
\centering
\includegraphics[trim=0 0 0 0,clip,width=\linewidth]{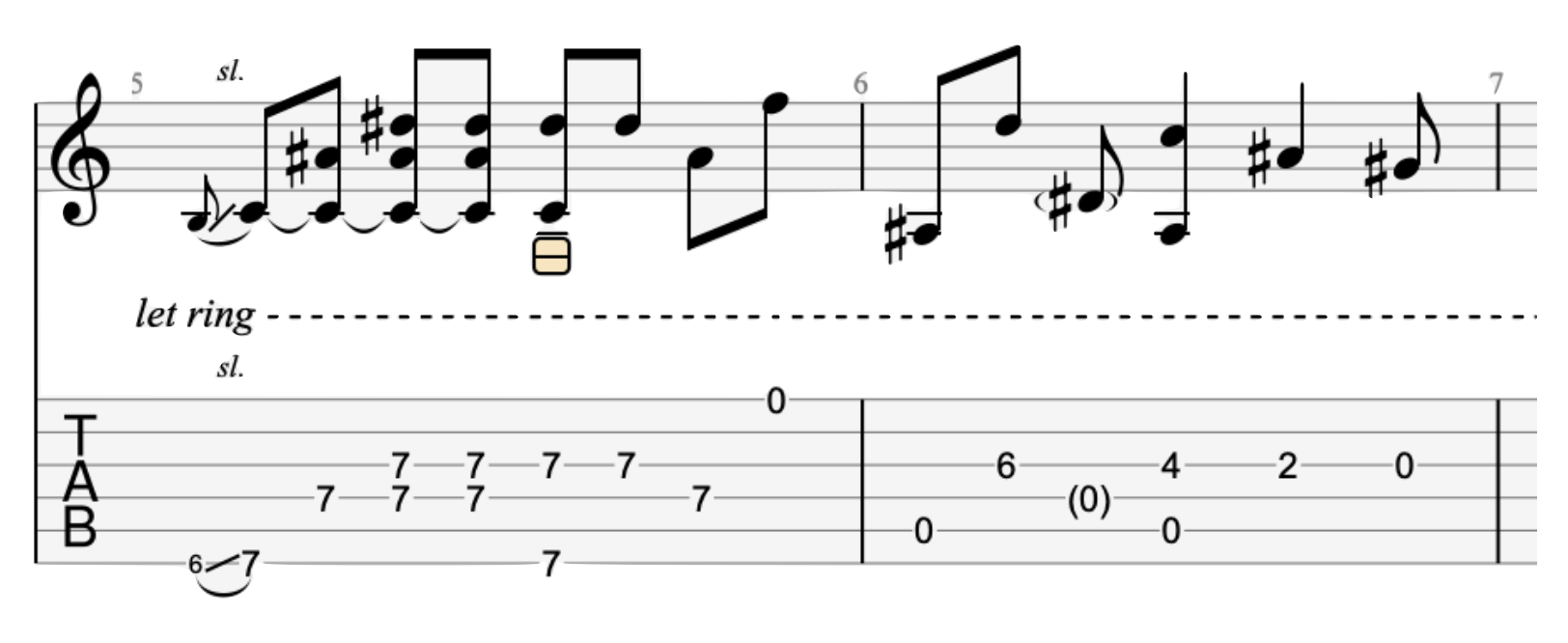}
\caption{An example of fingerstyle guitar tab composed by human, along with the corresponding staff notation.}
\label{fig:tab}
%\vspace{-3mm}
\end{figure}

There are, however, many other forms of musical notation that are quite different from the staff notation assumed by keyboard music. 
%other digital symbolic representations of music that are quite different from MIDI. 
For example, the tabulature, or ``tab'' for short, is a notation format that indicates instrument fingering rather than musical pitches.  It is common for fretted stringed instruments such as the guitar and ukulele, and free reed aerophones such as the harmonica. %Tablature was common during the Renaissance and Baroque eras, and is commonly used today in notating many forms of music.
It makes more sense for people playing such instruments to read the tabs, as they suggest how to move the fingers.

As shown in Figure \ref{fig:tab}, a tab contains information such as the fingering configuration on the fretboard (six strings for the case of the guitar) as well as usage of the left-hand or right-hand playing techniques. Such information is usually missing in the corresponding staff notation and  MIDI files. Learning to automatically compose guitar music directly from MIDI files, though possible, has the limitation of ignoring the way people play these instruments. However, to our best knowledge, little has been done to use tabs to train a deep generative model.

To investigate the applicability of modern deep learning architectures for composing tabs, we compile a new dataset of 333 TAB files of ``fingerstyle guitar'' (including originally fingerstyle guitar music and  fingerstyle adaptation) \cite{fingerstyle},
and modify the data representation of the Music Transformer \cite{huang2018music} to make the extended model learn to compose guitar tabs. 
%For convenience, we refer to the resulting model as a ``Guitar Transformer.'' 
With this model, we aim to answer three research questions (RQs):
\begin{itemize}
    \item Whether the neural network learns to generate not only the note sequences but also the fingering of the notes to be played on a fretboard, from reading only the tabs (instead of, for example, watching videos demonstrating how people play the guitar)? 
    \item Whether the neural network generates compositions with coherent ``groove,'' or the use of rhythmic patterns over time \cite{dixonEtAl04ismir,peeters05ismir,thompsonDM14}? 
    %We find this question intriguing,
    It is generally assumed that the layers of a neural network learn abstractions of data on their own to perform the intended task, e.g., to predict the next events given the history.
    %in a neural sequence model such as the Transformer \cite{vaswani2017attention}.
    However, in music, groove is usually indirectly implied according to the arrangement of notes along the time axis, instead of explicitly specified in either a MIDI or TAB file.
    Therefore, it remains to be studied whether the model can do better if it has to explicitly handle bar-level \texttt{GROOVING} events, inserted into the training data as a high-level information in some way, or if such a modification is not needed. This is in particular relevant in the context of fingerstyle composition, as in fingerstyle a guitarist has to take care of the melody, chord comping, bass line and rhythm simultaneously \cite{fingerstyle}.
    \item Finally, how the  compositions generated by the neural network compare with human-composed guitar tabs, when both rendered into audio waveforms and presented to human listeners? This gives us a direct evaluation of the effectiveness of the neural network in modeling guitar music.
\end{itemize}

We provide audio rendition of examples of the generated tabs (using a %commercial 
guitar synthesizer of a DAW called Ample Sound  \cite{amplesound}) 
at \url{https://ss12f32v.github.io/Guitar-Transformer-Demo/}, along with a video recording of a  guitarist playing a generated tab.

In what follows, we review some related work in Section \ref{sec:page_size}, and then present the tab dataset in Section \ref{sec:db}. After that, we describe in Section \ref{sec:method} the methodology for modeling and learning to compose guitar tabs. We present the result of objective and subjective evaluations addressing the aforementioned research questions in Section \ref{sec:exp}. %, and finally conclude the paper in Section \ref{sec:conclusion}.

\section{Related work}
\label{sec:page_size}

%While our literature survey is by no means comprehensive, we intend to highlight in this section a selected set of work that are relevant to our work.
%We highlight in this section a selected set of existing work that are relevant to our work.

\subsection{Guitar-related Research in MIR }

%Over the recent years the research about piano and keyboard type instrument had benefited from the progress of deep learning. Moreover, the data format for storing the instrument played by keyboard which we called 'MIDI' can easily keep the whole information of the playing. Despite the music information research (MIR) progress is rapidly being developed. 
%The MIR research in string instruments especially in guitar still remains challenging due to the data quantity and the format of the storage. 

In the music information retrieval (MIR) community, research concerning guitar is often related to automatic guitar transcription \cite{barbancho10taslp,Hrybyk2010CombinedAA,BurletEtAl_2013_RoboGuitTablTran,Yazawa13icassp,humphrey14icassp,christian14dafx,abeber17taslp,gorlow16arxiv,rodrguez18fma,Michelson2018AutomaticGT} and playing technique detection \cite{abesser2010icassp,su14ismir,chen15ismir,su19tismir}.
%For example, Kehling \emph{et al.} \cite{christian14dafx} developed an hybrid system to gradually detect the note onset, multi-$f_{0}$, offset, as well as string, plucking, expression style of electric guitar performances. 
%%In addition to guitar, \cite{abeber17taslp} apply a similar method on detecting note relevant information(i.e, onset, $f_{0}$, offset) and instrument relevant information(i.e, playing techniques, fretboard position) in bass. 
For example, Su \emph{et al.} \cite{su19tismir} built a convolution neural network (CNN) model for detecting the playing techniques associated with the string-pressing hand, and incorporated that for transcribing audio recordings of unaccompanied electric guitar performances.
Rodr{\'\i}guez \emph{et al.} \cite{rodrguez18fma} presented a model for transcribing Flamenco guitar falsetas, and Abe{\ss}er and Schuller \cite{abeber17taslp} dealt with the transcription of solo bass guitar recordings.
We note that, while automatic transcription concerns with recovering the tab underlying an audio guitar performance, our work deals with automatic composition of original guitar tabs in the symbolic domain, and therefore does not consider audio signals.

%automatic generation of the fingering arrangement given 
As there are multiple fret positions to play the same note on a guitar, it may not  be easy for a novice guitar learner to play a guitar song without the corresponding tab. Automatic suggestion of the fingering given a human-made ``lead sheet,'' a symbolic format that specifies the melody  and chord sequence but not their fingering, 
%a.k.a., tab fingering arrangement,
has therefore been a subject of research. Existing work has explored the use of hidden Markov models, genetic algorithm, and neural networks to predict the fingering by examining its playing difficulty for a guitarist, viewing the task as an optimal path finding problem \cite{tuohy2006guitar, sayegh1989fingering, mistler2017generating,arigaFG17ismir}. 
While such prior arts can be considered as performing a MIDI-to-TAB conversion, our work aims to model TABs directly.

Xi \emph{et al.} developed the GuitarSet \cite{xi2018guitarset},
%\footnote{\url{https://github.com/marl/guitarset/}}
a set of 360 audio recordings of a guitar equipped with the hexaphonic pickup. The special pickup is able to capture the sound from each string individually, making it possible for a model to learn to perform  multipitch estimation and tabulature fingering arrangement at the same time. Using the dataset, 
Wiggins and Kim \cite{wiggins2019guitar} built such a model with CNN, achieving 0.83 F-score (i.e., the harmonic average of precision and recall) for multipitch estimation, and 0.90 for identifying the string-fret combinations of the notes.
%, dubbed tabCNN, 
%detect the correct pitch by Pyin with several post-process methods to align with correspond audio. In automatic transcription, tabCNN use the previous mentioned dataset to apply CNN-based model to detect pitch along with the string and fret information. 
While the dataset is relevant for guitar transcription, its recordings are all around 12--16 bars in length only, which seems to be too short for deep generative modeling.
%building a language model involved in deep generative networks.
%training automatic tab composition models. 

McVicar \emph{et al.} \cite{mcvicar14icsp,  mcvicar14icmc,mcvicar15taslp} used to build sophisticated probabilistic systems to algorithmically compose rhythm and lead guitar tabs from an input %downbeat-synchronized
chord and key sequence. Our work differs from theirs in that we aim to build a general-purpose tab composition model using modern deep generative networks. %deep learning methods and that our model does not take chord or key inputs.
An extra complexity of our work is that we experiment with fingerstyle guitar, a type of  performance that can be accomplished by a single guitarist.

%\subsection{Automatic Music Composition}

%Over the past few years, the method of controllable music generation which based on language-model are discussed by several approaches. In natural language processing (NLP) domain, CTRL~\cite{keskar2019ctrl} extract condition codes from the raw text which still reserve the unsupervised training setting. The control code provide explicit prior information to the model which enable the generated text being controlled. 

\subsection{Transformer Models for Automatic Composition}

The Transformer \cite{vaswani2017attention} is a deep learning model that is designed to handle ordered sequences of data, such as natural language. 
It models a word sequence $(w_1, w_2, \dots w_T)$ seen in the training data by factorizing the joint probability into a product of conditionals, namely,
%\begin{equation}
$P(w_1) \cdot P(w_2|w_1) \cdot \dots \cdot P(w_T|w_1, \dots , w_{T-1}) \,.$
%\end{equation}
%P(w_1) \cdot P(w_2|w_1) \cdot \dots \cdot P(w_n|w_1, \dots , w_{n−1}) \dots \,.
During the training process, the model optimizes its parameters so as to correctly predict the next word $w_t$ given its preceding history $(w_1, w_2, \dots w_{t-1})$, for each position $t$ in a sequence.
%each sequence available in the training data. 

Following some recent work on recurrent neural network (RNN)-based automatic music composition \cite{melodyRNN16,oore2018time},  Huang \emph{et al.} \cite{huang2018music} viewed music as a language and for the first time employed the Transformer architecture for modeling music. Given a collection of MIDI performances, they converted each MIDI file to a time-ordered sequence of musical ``events,'' so as to model the joint probability of \emph{events} as if they are \emph{words} in natural language (see Section \ref{sec:method:event:MIDI} for details of such events).
The Transformer with relative attention was shown to greatly outperform an RNN-based model, called PerformanceRNN \cite{oore2018time}, in a subjective listening test \cite{huang2018music}, inspiring the use of Transformer-like architectures, such as Transformer or Transformer-XL \cite{dai2019transformer}, in follow-up research  \cite{donahue2019lakhnes,musenet,choi2019encoding,huang2020pop,wu20ismir}.\footnote{We note that it is debatable whether music and language are related. We therefore envision that some other new architectures people will come up with in the future might do a much better job than Transformers in modeling music. This is, however, beyond the scope of the current work.}

%A Transformer-based autoencoder proposed by \cite{choi2019encoding} used a melody encoder to leverage an auxiliary melody information into transformer decoder and separate the melody and the aspects of performance style. For multi-instrument generation, MuseNet~\cite{musenet} add a specify notation on every note related token (i.e., the number of token will be pitch number multiply instrument number) to achieve multi instrument generation. Furthermore, they add the composer and style token at the front of the input to indicate the global song information. 

%%from donahue:
%To model the event sequences outlined in the last section, we adopt a language modeling factorization. This factorization is convenient because it allows for a simple left-to-right algorithm for generating music: sampling from the distribution estimated by the model at each timestep (conditioned on previous outputs). The goal of our optimization procedure is to find a model configuration which maximizes the likelihood of the real event sequences. Motivated by the strong results for piano music generation from the recent Music Transformer [1] approach, we also adopt a Transformer [27] architecture.

There are lots of approaches to automatic music composition, deep learning- and non-deep learning based included \cite{Papadopoulos99aimethods,fernandez13jair,briot19book}. We choose to consider only the Transformer architecture here, to study whether we can translate its strong result in modeling MIDIs to modeling TABs.

\begin{table}
  \begin{center}
    \begin{tabular}{l|rrrr}
    \toprule
       & \multirow{2}{*}{\textbf{\# tabs}} & \multirow{2}{*}{\textbf{\# bars}} & \textbf{ \# bars } & \textbf{ \# events } \\
       & & & \textbf{per tab} & \textbf{per tab} \\
      \midrule
      training       & 303  & 24,381 & 80$\pm$41 & 5,394$\pm$3,116 \\
      validation    & 30  & 2,593  & 74$\pm$35 &  5,244$\pm$3,183 \\
      \bottomrule
    \end{tabular}
    \caption{Statistics of the dataset; the last two columns show the mean and standard deviation values across each set. Please see Table \ref{tab:vocab} for definitions of the events.}
    \vspace{-3mm}
    \label{tab:numbers}
  \end{center}
\end{table}

\section{Fingerstyle Guitar Tab Dataset}
\label{sec:db}

There have been some large-scale MIDI datasets out there, such as the Lakh MIDI dataset \cite{raffel16ismir} and BitMidi \cite{bitmidi}. 
The former, for example, contains 176,581 unique MIDI files of \emph{full songs}. In contrast, existing datasets of tabs are usually smaller and shorter, as they are mainly designed for learning the mapping between tabs and audio (i.e., for transcription research), rather than for generative modeling of the structure of tabs.
The tabs in the GuitarSet \cite{xi2018guitarset}, for example, are performances of \emph{short excerpts of songs}, typically 12--16 bars in length, which are not long.

For the purpose of this research, we compile a guitar tab dataset on our own, focusing on the specific genre of fingerstyle guitar. Specifically, we collect digital TABs of \emph{full songs}, to facilitate language modeling of guitar tabs. We go through all the collected TABs one-by-one and filter out those that are of low quality (e.g., with wrong fingering, obvious annotation errors), or are not fingerstyle (e.g., have more than one tracks).
We also discard TABs that are not in standard tuning, to avoid inconsistent mapping between notes and fingering.
As shown in Table \ref{tab:numbers}, this leads to a collection of 333 TABs, each with around 80 bars. This includes TABs of famous professional fingerstyle players such as Tommy Emmanuel and Sungha Jung. 
All the TABs are in 4/4 time signature, and they can be in various keys.
%Therefore, we decide to scrap the tablature data from the open released database and focus on specific playing style. We obtained digital guitar tabs for a selection of guitarists from "https://guitarprotabs.org/" and "http://gtptabs.com/".
We reserve 30 TABs for validation and performance evaluation, and use the rest for training.

%In this section, we introduce the method of how we collect the tablature data, pre-processing method and introducing fingerstyle playing .

%\subsection{Collection of tabulature data}\label{subsec:body}
%Since we are interesting in guitar tab generation. In previous music generation work, most of them are focusing are piano generation. To the best of our knowledge, the only existed dataset which contains pitch and string well aligned is Guitarset~\cite{xi2018guitarset}. Even though, the data quantity of Guitarset still note enough for us to train an model to capture long sequence feature of guitar music, since the original usage of Guitarset is design for guitar transcription. Therefore, we decide to scrap the tablature data from the open released database and focus on specific playing style.
%We obtained digital guitar tabs for a selection of guitarists from "https://guitarprotabs.org/" and "http://gtptabs.com/".

Please note that, similar to the MIDI files available in Lakh MIDI \cite{raffel16ismir}, the TAB files we collect do not contain \emph{performance} information such as expressive variations in dynamics (i.e., note velocity) and micro-timing \cite{lerch19ismir,oore2018time}. To increase velocity variation, we use Ample Sound \cite{amplesound} to add velocity to each note by its humanization feature. We do not deal with  micro-timing in this work.

%\subsection{Preprocess}\label{subsec:body}

%After we collect a amounts of tablature, we first filter our these type of data to make sure our pitch and playing position mapping are consistent. Then we discard the multi-guitar tabs to avoid the problem that simply combine multi-tracks to single tracks(i.e, combine results may not be played human.) A total number of 303 pieces of standard tuning tablature data played by several famous professional (e.g.,Sungha Jung, Tommy Emmanuel) fingerstyle player are collected finally. 

\subsection{Fingerstyle}
It is interesting to focus on only fingerstyle guitar in the context of this work, as we opt for validating the effectiveness of Transformers for single-track TABs first, before moving to modeling multi-track performances that involve at least a guitar (e.g., a rock song). We give a brief introduction of fingerstyle guitar below.

Fingerstyle \cite{fingerstyle} is at first a term that describes using fingertips or fingernails to pluck the strings to play the guitar. Nowadays, the term is often used to describe an arrangement method to blend multiple parts of musical elements or tracks, which are initially played by several instruments, into the composition of one guitar track. 
Therefore, a guitarist playing fingerstyle has to simultaneously take care of the melody line, bass line, chord comping and the rhythmic groove.
Groove, in particular, is important in fingerstyle, as it is now only possible to work on the rhythmic flow of music with a single guitar and the use of the two hands.
%Fingerstyle arrangement also needs to consider the grooving by the note and percussion technique combination, since the grooving can only be controlled by a single guitar and the use of two hands now. 
%As we do not plan to discuss multi-instrument playing (i.e, dual guitar) in this work. To make the result can be individually demonstrate, fingerstyle is the best choice and also the suitable data for discussing grooving.
We hence pay special attention to groove modeling in this work (see Section \ref{subsec:grooving}).

% For the aligned string information, we transform each string note to a \texttt{STRING} and \texttt{FRET} event token and concatenate after the pitch note related event token. By combining the pitch note and string information, the whole information in the tab can be embedded in to the model input.

\begin{table}
  \begin{center}
    %\scalebox{0.9}{
    \begin{tabularx}{\linewidth}{l|l}
    \toprule
    \textbf{category/type} & \textbf{description} \\
    \midrule
    \texttt{NOTE-ON} &  45 different pitches (E2--C6)\\
    \texttt{NOTE-DURATION} & multiples of the 32th note (1--64)\\
    \texttt{NOTE-VELOCITY} & note velocity as 32 levels (1--32)\\
    \texttt{POSITION} & temporal position within a bar;   \\
    & multiples of the 16th note (1--16)\\
    \texttt{BAR} & marker of the transition of bars\\
    \midrule
    \texttt{STRING} & 6 strings on a tab\\
    \texttt{FRET} & 20 fret positions per string\\
    \texttt{TECHNIQUE} & 5 playing techniques: slap, press \\
    &  upstroke, downstroke, and hit-top \\
    \texttt{GROOVING} & 32 grooving patterns\\
    \bottomrule
    \end{tabularx}
    %}
    \caption{The list of events adopted for representing a tab as an event sequence. The first five are adapted from \cite{huang2018music,huang2020pop}, whereas the last four are tab-specific and are new. We have in total 45+64+32+16+1+6+20+5+32$=$231 unique events.}
    \label{tab:vocab}
  \end{center}
\end{table}

\section{Modeling Guitar Tabs}
\label{sec:method}
%New Event-based representation of string type music

In this section, we elaborate how we design an event representation for modeling guitar tabs, or more generally tabs of instruments played by string strumming.

\begin{figure}[]
\centering
\includegraphics[width=\linewidth]{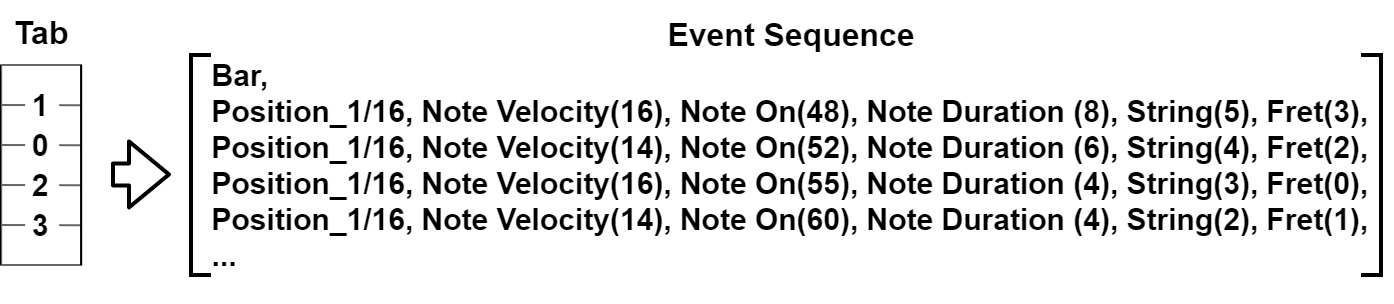}
\caption{An example of the result of  ``TAB-to-event'' conversion needed for modeling a tab as a sequence. Here, we show the resultant event representation of a C chord.}
\label{fig:example_event}
\end{figure}

%\subsection{Problem Formulation}
%\section{Model}\label{sec:grooving}

% We also set the memory context length to 512 in able to capture longer information to generate a high consistency result.(512 events approximately equal to 100 notes on the tab).

%\subsection{Event Representation for Tabs}
%\label{sec:method:event}

\subsection{Event Representation for MIDIs: A Quick Recap}
\label{sec:method:event:MIDI}

In representing MIDIs as a sequence of ``events,'' Huang \emph{et al.} \cite{huang2020pop} considered, amongst others, the following event tokens. Each  note is represented by a triplet of \texttt{NOTE-ON}, \texttt{NOTE-DURATION}, and \texttt{NOTE-VELOCITY} events, representing the MIDI note number, quantized duration as an integer multiple of a \emph{minimum duration}, and discrete level of note dynamics, respectively. 
The minimum duration is set to the 32th note. 
% the duration of notes in a MIDI score is quantized accordingly. 
%For the basic duration unit, they quantize the durations to multiples of the 
% i.e., the shortest duration of a note is the 32th note.
The onset time of the notes, on the other hand, is marked (again after quantization) on a time grid with a specific \emph{resolution}, which is set to the 16th note as in \cite{huang2018music}. Specifically, to place the notes over the 16-th note time grid, they use a combination of  \texttt{POSITION} 
and \texttt{BAR} events, indicating respectively the position of a note onset within a bar, among the 16 possible locations, and the beginning of a new bar as the music unfolds over time. This event representation has been shown effective in modeling pop piano \cite{huang2020pop}.
We note that the time grid outlined with this combination of \texttt{POSITION} and \texttt{BAR} events can also contribute to modeling the rhythm of fingerstyle guitar.

%Since we are focusing on generating fingerstyle guitar music, the repeating bassline and rhythmic structure are one of the main point to distinguish between the generated music and realistic music. In order to capture the above two characteristics, we adopt the \texttt{POSITION} and \texttt{BAR} setting same as REMI to provide a clear boundary between each bar interval and also each note.

%\subsubsection{String and Fret}
%\label{subsec:body}
%As the MIDI is designed for the piano or keyboard instrument, it's hard to represent the string information such as the playing position of the current note. This is also the reason we use tablature data for the work. 

\subsection{Event Representation for Tabs}
\label{sec:method:event:TAB}

To represent TABs, we propose to add, on top of the aforementioned five types of events for MIDIs,\footnote{Huang \emph{et al.} \cite{huang2020pop} actually considered the \texttt{Chord} and \texttt{Tempo} events additionally; we found these two types of event less useful in modeling tabs, according to preliminary experiments.} the following three new types of fingering-related events: \texttt{STRING}, \texttt{FRET},  \texttt{TECHNIQUE}, and a type of rhythm-related events: \texttt{GROOVING}. We introduce the first three below, and the last in the next subsection. Table \ref{tab:vocab} lists all the events considered, whereas
%in the proposed representation, and, 
Figure \ref{fig:example_event} gives an example of how we represent a C chord with such an event representation.

We use the first 20 frets of the 6 strings in the collected TABs, i.e., each string can play 20 notes. The pitch range of the strings overlaps, so a guitarist can play the same pitch on different strings, with moderate but non-negligible difference in timbre. The fingering of the notes also affects playability \cite{Yazawa13icassp}. In standard tuning, the strings can play 45 different pitches, from E2 to C6.
%which is often assumed, the open-string pitches of the strings are E, A, D, G, B, and E, respectively, from lowest (low E2) to highest (high E4). Therefore, there are in total 45 unique pitches.
%Standard tuning is used by most guitarists, and frequently used tunings can be understood as variations on standard tuning.

In our implementation, we adopt the straightforward approach to account for the various possible playing positions of the notes---to add \texttt{STRING} and \texttt{FRET} tokens right after the \texttt{NOTE-ON} tokens in the event sequence representing a tab. We note that the \texttt{FRET} tokens are actually \emph{redundant}, in that the combination of \texttt{NOTE-ON} and \texttt{STRING} alone is sufficient to determine the fret position to use. However, in pilot studies we found the inclusion of \texttt{FRET} makes the model converges faster at the training time.

Specifically, instead of a 3-tuple representation of a note as the case in MIDIs, we use a 5-tuple note representation that consists of successive tokens of \texttt{NOTE-VELOCITY}, \texttt{NOTE-ON}, \texttt{NOTE-DURATION}, \texttt{STRING} and \texttt{FRET} for TABs. As such five tokens always occur one after another in the training sequences, it is easy for a Transformer not to miss any of them when generating a new \texttt{NOTE-ON} event at the inference time, according to our empirical observation of the behavior of the Transformers.

However, as we do not impose constraints on the association between \texttt{NOTE-ON} and \texttt{STRING}, it remains to be studied whether a Transformer can learn to compose tabs with reasonable note-string combinations.
%on its own. 
This is the subject of the \textbf{1st RQ} outlined in Section \ref{sec:introduction}.

%after the note related tokens. It clearly indicate the string and the fret position.  In this way, we can easily to inform the model that the \texttt{STRING} and \texttt{FRET} tokens are belong to the prior note, and also benefit from the flexible combination (since one note may have different place to be played in guitar).

%After we filter out the non-suitable data, we transform our tab to MIDI that each pitch note are accompanied with a string note to indicate the string originally notated on the tab. Thanks to our source are from tab format which has a clear bar notation and correct note position, we can easily apply REMI \cite{huang2020pop} representation on our pitch note by also quantize every bar to a 16 grid. 

As for the \texttt{TECHNIQUE}s, we consider the following five right-hand techniques: slap, press, upstroke, downstroke, and hit-top, which account for $\sim$1\% of the events in our training set. The inclusion of other techniques, such as sliding and bending, is left as a future work.

%\subsubsection{Note related information}\label{subsec:body}

Similar to \cite{huang2018music,huang2020pop}, we consider the 16th note as the resolution of onset times, which is okay for 4/4 time signature. Increasing the  resolution further to avoid quantization errors and to enhance expressivity is also left to the future.

%\yh{Maybe say something on velocity here...?}

%Previous work are focusing on piano generation, which use MIDI score as a training data. Based on REMI, We proposed a new representation for string type instrument, especially for guitar, which has multi ambiguous playing choices for same pitch. 

\begin{figure}[]
\centering
\includegraphics[width=.47\linewidth]{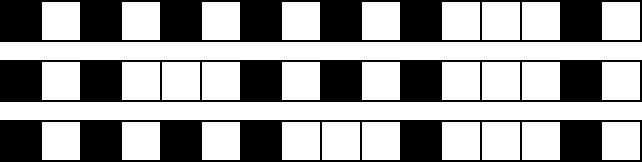} ~~~~
\includegraphics[width=.47\linewidth]{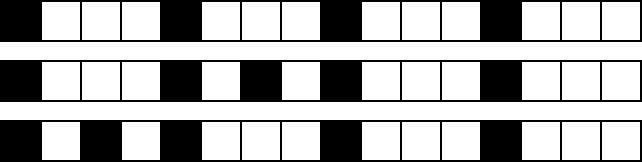}
\\ (a) ~~~~~~~~~~~~~~~~~~~~~~~~~~~~~~~~~~~~~~~~ (b)
\caption{Samples of 16-dim hard grooving patterns assigned to 2 different clusters (a), (b) by $k$means clustering.}
\label{fig:example_kmeans}
\end{figure}

%\subsection{Grooving Modeling}
%\label{sec:grooving}
%In this section, we  will first introduce the grooving and show several methods of how we represent a grooving.

%\subsection{Definition of Grooving}\label{subsec:grooving_def}

\subsection{Groove Modeling}
\label{subsec:grooving}

Groove can be in general considered as a rhythmic feeling of a changing or repeated pattern, or ``humans’ pleasurable urge to move their bodies rhythmically in response to music'' \cite{senn18groove}. 
%The term groove was firstly used in mid-to-late African-American styles such as R\&B. Nowadays, various types of grooving have been used across different genres. For example, in Jazz, we often say ``swing,'' which there is a ``flow'' in the music \cite{dittmarPM15}. 
Unlike the note-related or time-related events, groove is usually \emph{implicitly} implied as a result of the arrangement of note onsets over time, instead of be explicitly specified in either a MIDI or TAB file. Hence, it might be possible for a Transformer to learn to compose music with reasonable groove, without we \emph{explicitly} inform it what groove is. We refer to this baseline variant of our Transformer as the \textbf{no grooving} version, which considers all the events listed in Table \ref{tab:vocab} but \texttt{GROOVING}.

However, as a tab is now represented as a sequence of events, it is possible to add groove-related events to help the model make sense of this phenomenon.  Since our event representation has the \texttt{BAR} events to mark the bar lines, we can ask the model to learn to generate a ``bar-level'' \texttt{GROOVING} event right after a \texttt{BAR} event, before proceeding to generate the actual content of the bar. Whether such a groove-aware approach benefits the quality of the generated tabs is the subject of our \textbf{2nd RQ}.

To implement such an approach, we need to come up with 1) a bar-level grooving representation of symbolic music, and 2) a method to convert the grooving representation, which might be a vector, to a discrete event token.

%As our goal is to solve the discontinuous issue in prompt given generation scenario, we believe that the problem is that the transformer model using REMI representation only learn the sequential pattern of the notes but lack of the structure information between the each bar. To solve this problem, we design 5 methods to represent the grooving.

%Overall, grooving cannot be easily defined by a single meaning in music domain not to say in an algorithmic way. 
In this work, we represent groove by the \emph{occurrence of note onset} over the 16-th time grid, leading to the following four grooving representations of music.
%We investigate the following different ways to represent the grooving in this work:
\begin{itemize}
    \item \textbf{Hard grooving}: A 16-dim  binary vector marking the presence of (at least one) onset per each 16 positions of a bar.
    %We only care about the instant plucking sound in each position of the bar. Thus, we didn't care about the note-on quantity in this setting, instead we set the value of the grid that exist any note to 1 (e.g., if the player plays a chord in a designated time grid,  the value correspond designated time grid in the grooving sequence will be 1).
    A popular pattern in our dataset, for example, is  $[1,0,0,0,1,0,0,0,1,0,0,0,1,0,0,0]$, meaning onsets on beats only.
    \item  \textbf{Soft grooving}: A soft version that considers the number of onsets (but disregarding the velocity values) for each position, normalized by the maximum in the bar, leading to a 16-dim real-valued 
    %(in $[0,1]$) 
    vector. 
    %note quantity (e.g., difference between chord and arpeggios). the grooving sequence will be normalized to [0,1]. The higher value indicate the higher note density. 
    \item \textbf{Multi-resolution hard (or soft) grooving}: Variants of the last two that additionally consider corresponding down-sampled 8-dim and 4-dim vectors to emphasize the beats (e.g., counting only the onsets on beats), and then concatenate the vectors together, yielding a 28-dim vector (i.e., 16+8+4). 
    %In the two previous settings, they only consider the grooving of the bar which is quantized to 16 regions. To inform the model a hierarchical grooving resolution information, we extract a lower resolution grooving which is in 4 regions and 8 regions. The lower resolution grooving are also considered as if a note-on exist in the every 4th note or 8th note.
    
    %\item \textbf{Top 31 grooving sequence frequency representation}: As we quantize the grooving sequence to hard grooving representation. We are curious about:
    %\begin{itemize}
    %\item How many unique grooving sequence exist in the dataset ? 
    %\item Can we just only encode top \textit{N} frequency grooving sequence to represent grooving? 
    %\end{itemize}
    %To validate our assumption, we first transform the grooving sequence in hard grooving setting. Then, we calculate the occurrence frequency of each grooving sequence. We found that the most common grooving sequence are [1,0,1,0,1,0,1,0,1,0,1,0,1,0,1,0] which the occurrence frequency are 15.8\% (The total number of the bar are 26794). Thus, we decide only to encode the top 31 occurrence frequency by give each of them an special \textit{id} and the remains are aggregate to the same \textit{id}. 
\end{itemize}

% \subsubsection{Top 31 and other grooving}\label{subsec:grooving_hard}
% After we finish the hard grooving representation, we then make an analysis of the number of occurrences of each unique grooving sequence. We found that some general common grooving sequence dominate the number of occurrences in whole training data. The table show the top 31 frequency grooving sequence of whole data. 

% \begin{table}
%  \begin{center}
%  \begin{tabular}{|l|l|}
%   \hline
%   Grooving Sequence & Number of occurrences \\
%   \hline
%   1,0,1,0,1,0,1,0,1,0,1,0,1,0,1,0 &  4264 \\

%   1,0,0,0,1,0,0,0,1,0,0,0,1,0,0,0 &  1010 \\
  
%   1,0,0,0,1,0,1,0,1,0,1,0,1,0,1,0  &  841 \\
%   1,0,0,0,1,0,1,0,1,0,1,0,1,0,1,0  &  841 \\

%   \hline
  
%  \end{tabular}
% \end{center}
%  \caption{Table captions should be placed below the table.}
%  \label{tab:example}
% \end

% \begin{figure}[t]
% \centering
% \includegraphics[width=\linewidth]{figs/key_error_rate.png}
% \caption{Error rate vs MIDI number.}
% \label{fig:user_study1}
% \end{figure}

%\subsection{Modeling Grooving in Transformers}
%\label{subsec:grooving_model}

%We have mentioned that we design 5 grooving representation to encode the grooving information. Obviously, in a hard grooving setting, number of all the permutation of the grooving is $2^{16}$. How to incorporate such a high dimension information in to Transformer-XL is now a problem for us. 
To convert the aforementioned grooving patterns to events, a discretization is needed. Among various possible approaches, we experiment with the simplest idea of grouping the grooving patterns seen in the training set into a number of clusters. We can then use the ID of the cluster a grooving pattern is associated with for the \texttt{GROOVING} event of that grooving pattern.
%In GrooVAE \cite{gillick2019learning}, they use an RNN-based autoencoder with variational loss to learn the representation of each feature. Auto-encoder architecture also enable them to perform grooving transfer by changing the intermediate hidden vector from encoder to generate different offset and velocity. In this paper, as we not going to perform an transfer task. To tokenize our grooving sequence into our model,
For simplicity, we employ the classic $k$means algorithm \cite{hartigan1979kmeans} here, setting $k$ to 32. %, 8 grids use $k=16$, 4 grids use $k=4$ for the $k$means setting.  
Please see Figure \ref{fig:example_kmeans} for an example of the clustering result.

%\begin{table}[t]
%\centering
%%\setlength{\tabcolsep}{5pt} %% Default value: 6pt
%\begin{tabular}{l|c}
%\toprule
%Pitch & Error rate \\ 
%\midrule
%47   & 23.90\%  \\
%45  & 22.80\%\\
%44   & 18.57\%\\
%42  & 17.67\% \\
%49   & 16.38\%\\
%\bottomrule
%\end{tabular}
%\caption{Top 5 error rate of the pitch with wrong string.}
%\label{tab:pitch_error_rate}
%\end{table}

\subsection{Transformer-XL-based Architecture}
\label{sec:exp:implementation}

Following \cite{donahue2019lakhnes,huang2020pop}, we use the Transformer-XL \cite{dai2019transformer} for the architecture of our model. Unlike the Transformer used in \cite{huang2018music}, the Transformer-XL gains a longer receptive filed with a segment-level recurrence mechanism, thereby seeing further into the history and benefiting from the extended memory.
%and relative position embedding scheme to solve the time series problem in a longer context. 
%In \cite{huang2020pop}, they use Transformer-XL on pup music generation result in a better result on generating a higher context related music.
%We also adopt Transformer-XL in this work. 
We base our implementation on the open source code of \cite{huang2020pop}, adopting many of their settings.
For example, we also set the sequence length and recurrence length to 512 events, and use 12 self-attention layers and 8 attention heads. 
The model has in total $\sim$41M trainable parameters. The training process converges within 12 hours on a single NVIDIA V100 GPU, with batch size 32.

\section{Evaluation}
\label{sec:exp}

%We aim to answer the three RQs listed in Section \ref{sec:introduction} below.

% \begin{figure}[]
% \centering
% \includegraphics[width=\linewidth]{figs/string epoch.png}
% \caption{String matching error rate respect to epochs}
% \label{fig:user_study1}
% \end{figure}

% \centering
% \includegraphics[width=\linewidth]{figs/note_quantity.jpg}
% \caption{Note quantity in training dataset.}
% \label{fig:user_study1}
% \end{figure}

%\yohua {need to explain why error occurred in low pitch}
%\begin{figure}[]
%\centering
%\includegraphics[width=\linewidth]{figs/5-2-ErrorEvaluation_2.png}
%\caption{String relevant output probability from model of 3 specific pitches.}
%\label{fig:5-2}
%\end{figure}

% \begin{table}[t]
% \centering
% %\setlength{\tabcolsep}{5pt} % Default value: 6pt
% \begin{tabular}{l|c|c}
% \toprule
%  & k & \# of possible grooving sequence\\ 
% \midrule
% hard  & 32 & $2^{16}$ \\
% soft  & 32 & $2^{16}$\\
% hard-multi-16 & 32 & $2^{16}$\\
% hard-multi-8 & 16 & $2^{8}$ \\
% hard-multi-4 & 4 & $2^{16}$\\

% soft-multi-16 & 32 & $2^{16}$ \\
% soft-multi-8 & 16 & $2^{8}$ \\
% soft-multi-4 & 4 & $2^{16}$\\

% \bottomrule
% \end{tabular}
% \caption{Result of objective evaluation.}
% \label{tab:obj_result}
% \end{table}

\begin{figure}
    \centering
    \includegraphics[width=0.48\columnwidth]{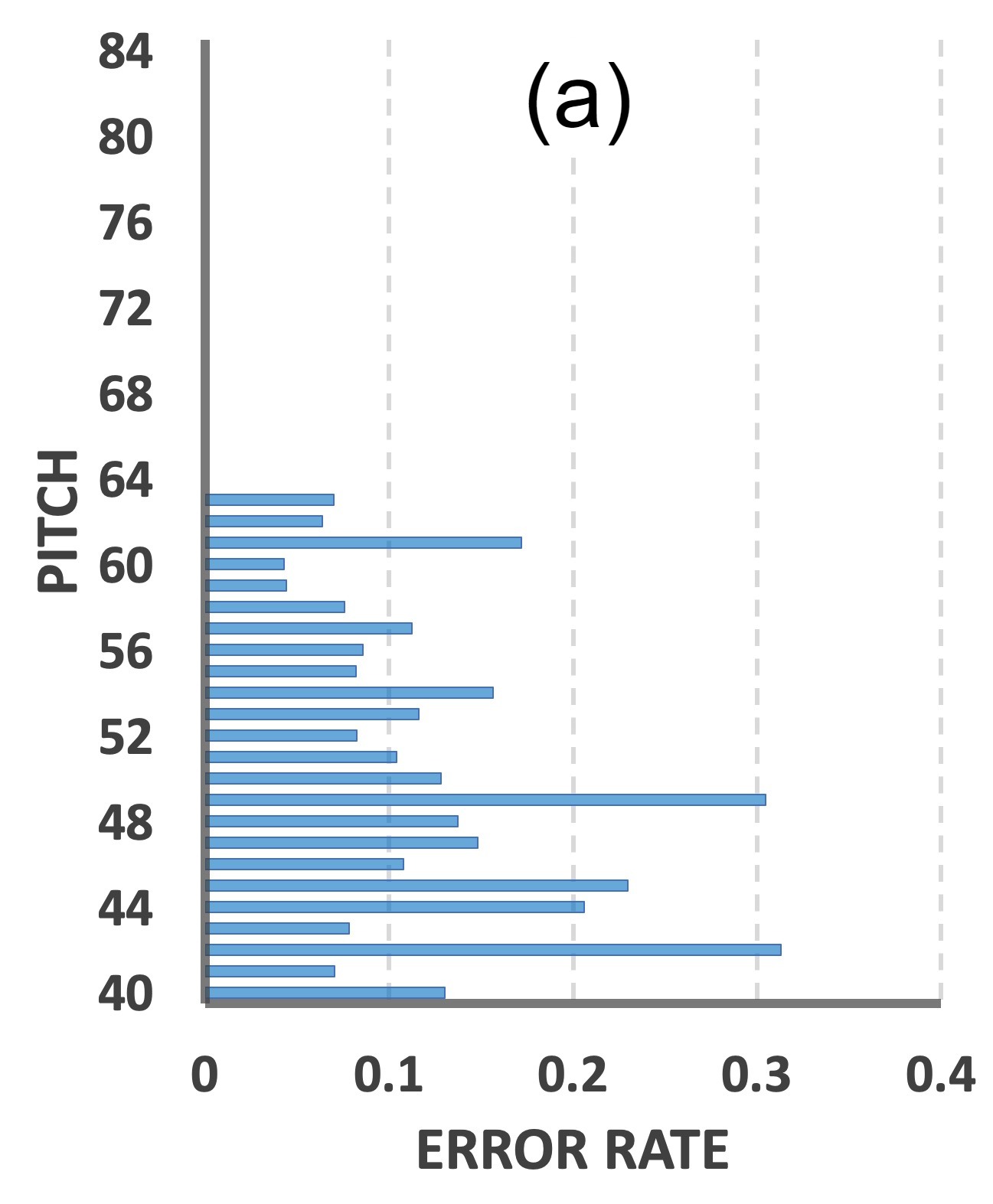}
    \includegraphics[width=0.48\columnwidth]{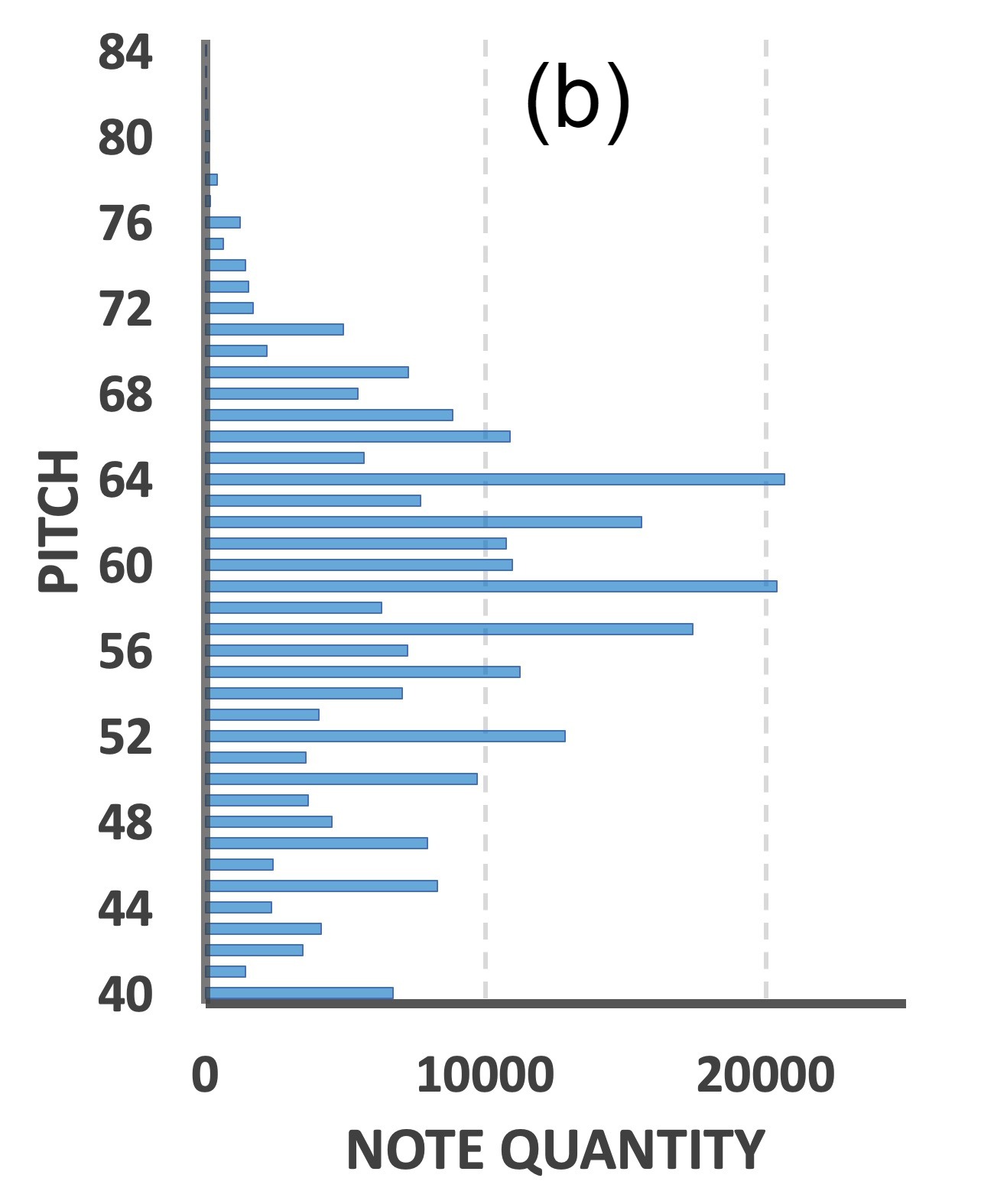} 
    \\
    % (a) ~~~~~~~~~~~~~~~~~~~~~~~~~~~~~~~~~~~~~~~~ (b)
    \caption{Distributions of (a) the error rate for each note in the string arrangement prediction of our model, and (b) the counts of each note in the training set.}
    \label{fig:result1}
\end{figure}

\begin{table}[t]
\centering
\scalebox{0.88}{
\begin{tabular}{l|cccccc}
\toprule
 & \multicolumn{6}{c}{\textbf{string}  (high-pitched $\leftrightarrow$ low-pitched) } \\
 & \textbf{1st}  & \textbf{2nd} & \textbf{3rd} & \textbf{4th} & \textbf{5th} & \textbf{6th}  \\ 
\midrule
(a) accuracy   & 100\% & 99\% &  97\% & 94\%  & 91\%  & 90\% \\
%Error rate   & 9.57\% & 8.60\% &  5.82\% & 2.74\%  & 0.77\%  & 0.00\% \\
\midrule
(b) pitch 42 & $\sim$0\% & $\sim$0\% & 10\% & $\sim$0\% & 27\% & 63\%
\\
(c) pitch 57 & $\sim$0\% & 6\% & 65\% & 26\% & $\sim$0\% & $\sim$0\%
\\
(d) pitch 69 & 85\% & 14\% & $\sim$0\% & $\sim$0\% & $\sim$0\% & $\sim$0\%
\\
\bottomrule
\end{tabular}
}
\caption{(a) The average accuracy of our model in associating each \texttt{STRING} with a \texttt{NOTE-ON}, broken down by string; (b--d) The string-relevant output probability estimated by our model for three different pitches.}
\label{tab:string_error_rate}
\end{table}

\begin{table}[t]
\centering
\begin{tabular}{l|cc|cc}
\toprule
 & \multicolumn{2}{c}{\textbf{Hard accuracy} $\uparrow$}  & \multicolumn{2}{c}{\textbf{Soft distance}  $\downarrow$} \\ 
& mean & max & mean & min \\ 
\midrule
hard grooving & 76.2\% & 82.4\% & 56.3 & 44.6 \\
soft grooving & 76.9\%& 83.0\% & \textbf{56.2} & \textbf{43.7} \\
multi-hard & \textbf{79.0\%} & \textbf{85.7\%} & 57.8 & 44.3 \\
multi-soft & 74.6\% & 81.1\% & 64.7 & 52.9 \\
%top32 
\midrule
no grooving & 70.0\% & 80.1\% & 58.6 & 47.7\\
\midrule
training data  & \textbf{82.1\%} & \textbf{89.5\%} & \textbf{43.8} & \textbf{28.6} \\
random & 64.9\% & 71.3\% & 70.6 & 59.6 \\
\bottomrule
\end{tabular}
\caption{Objective evaluation on groove coherence.}
\label{tab:obj_result}
\end{table}
% https://docs.google.com/spreadsheets/d/1CqWesKHqbID2m4lv8RtU-S45lCYBRROwtxRKNggu4xk/edit#gid=0

\begin{figure*}[!ht]
\centering
\includegraphics[width=.8\textwidth]{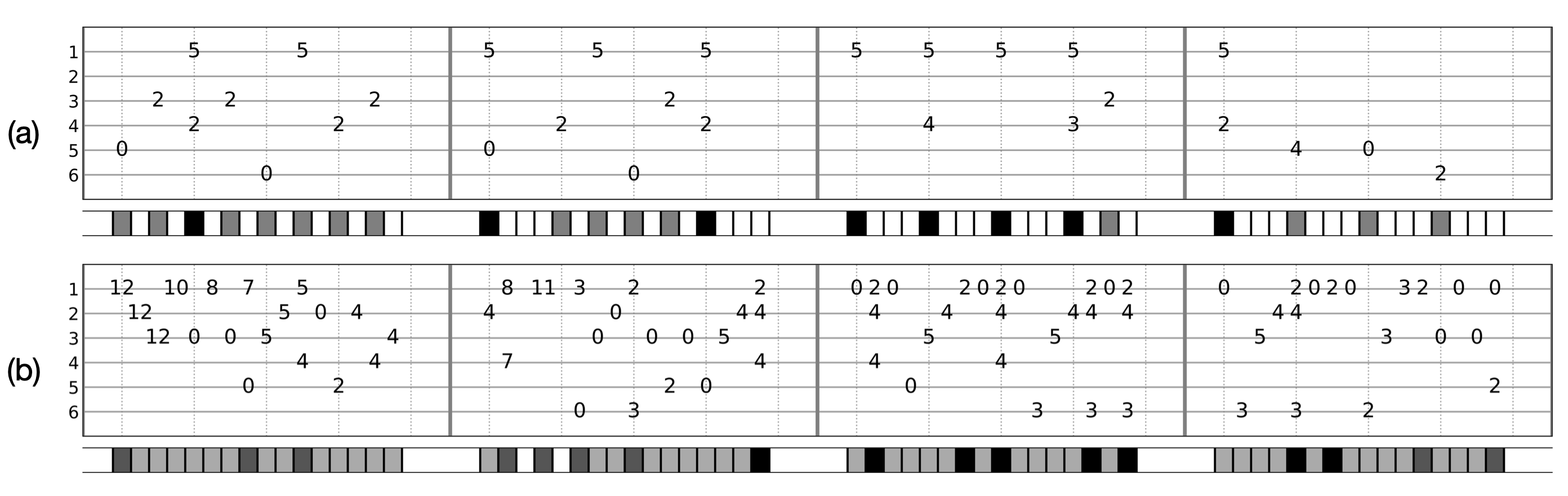}
\caption{Segments of 2 tabs randomly generated by the hard grooving model; below each tab---the soft grooving patterns.}
\label{fig:generatd_result_to_tab}
\end{figure*}

\subsection{Experiment 1: On Fingering}
\label{sec:exp1}

The 1st RQ explores how a Transformer learns the association between notes and fingering, without human-assigned prior knowledge/constraints on the association. For simplicity, we use the \textbf{no grooving} variant of our model here.  

A straightforward approach to address this RQ is to let the model generates randomly a large number of event sequences (i.e., compositions) and examine how often it generates a plausible \texttt{STRING} event after a \texttt{NOTE-ON} event.
Table \ref{tab:string_error_rate}(a) shows the average note-string association accuracy calculated from 50 generated 16-bar tabs, broken down into six values according to \texttt{STRING}.
To our mild disappointment, the accuracy, though generally high, is not perfect.  This indicates that some post-processing is still needed to ensure the note-string association is correct.

As Table \ref{tab:string_error_rate}(a) shows larger errors toward the 6th string, we also examine how the errors distribute over the pitches.
Interestingly, Figure \ref{fig:result1}(a) shows that the model makes mistakes only in the low end; the fingering prediction is good for pitches (i.e., MIDI numbers) from 64 to 84. 

It is hard to find out why exactly this is the case, but we present two more observations here. First, we plot in Figure \ref{fig:result1}(b) the popularity of these pitches in the training set. The Pearson correlation coefficient between the note quantity and the error rate is weak, at 0.299, suggesting that this may not be due to the sparseness of the low-pitched notes.
Second, we show in Table \ref{tab:string_error_rate}(b)--(d) the note-string association output probability estimated by our model for three different pitches. Interestingly, it seems the model has the tendency to use neighboring strings for each pitch. For example, pitch 42 is actually a bass note playable on the 6th string, and it erroneously ``leaks'' mostly to the 5th string.

\begin{figure}[t]
\centering
\includegraphics[trim=1.5 0 0 1,clip,width=.85\linewidth]{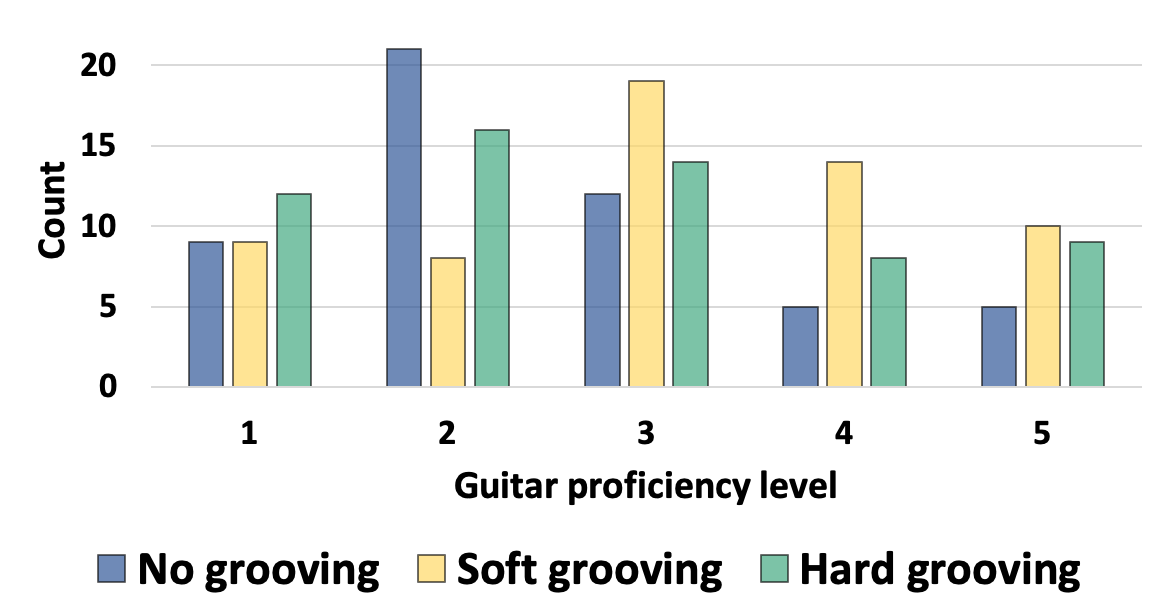}
\caption{Result of the first user study asking subjects to choose the best among the three continuations generated by different models, with or without \texttt{GROOVING}, given a man-made prompt. The result is broken down according to the self-report guitar proficiency level of the subjects.}
\label{fig:user_study1}
\end{figure}

%\begin{figure}[t]
%\centering
%\includegraphics[width=\linewidth]{figs/user_study3.png}
%\caption{Result of optional question from high level guitar proficiency participants in first user study.}
%\label{fig:user_study3}
%\end{figure}

\subsection{Experiment 2: On Groove}
\label{sec:exp1}

Figure \ref{fig:generatd_result_to_tab} gives two examples of tabs generated by the hard grooving model. It seems the grooving is consistent across time in each tab. But, how good it is?  

The 2nd RQ tests whether the added \texttt{GROOVING} events help a Transformer compose tabs with better rhythmic coherence.  We therefore intend to compare the performance of models trained with or without \texttt{GROOVING} for generating ``continuations'' of a given ``prompt.''

%Since the measurement of the music still remains a open question and hard to evaluate the quality of generated music. In this paper, we proposed two new methods to evaluate the consistency and coherence between the given prompt and continuous generated music. 

We consider both objective and subjective evaluations here. For the former, we compare the models trained with \texttt{GROOVING} events obtained with each of the four vector-quantized grooving representations described in Section \ref{subsec:grooving}. We ask the models to generate 16-bar continuations following the first 4 bars of the 30 tabs in the validation set. The performance of the models is compared against that of the `no-grooving' baseline, the `real' continuations (of these 30 tabs), and a `random' baseline that picks the next 16 bars from another tab at random from the validation set. The last two are meant to set the high-end and low-end performances, respectively. For fair comparison, we also project the note onsets of the validation data onto the 16th-note grid underlying our training data.

We consider the following two simple objective metrics:
\begin{itemize}
    \item \textbf{Hard accuracy}: Given the hard grooving patterns $\mathbf{X}=(\mathbf{x}_1, \dots, \mathbf{x}_N)$ of the prompt, and those of the continuation $\mathbf{Y}=(\mathbf{y}_1, \dots, \mathbf{y}_M)$, where both $\mathbf{x}_i$ and $\mathbf{y}_j$ are in $\{0,1\}^{K}$, $N=4$, $M=16$, $K=16$, we compare the similarity between $\mathbf{X}$ and $\mathbf{Y}$ by
    \begin{equation}
    \text{mean}_{i=(1,\dots,N)}
    \frac{1}{MK} \sum_{j=1}^{M}   \sum_{k=1}^{K} \text{XNOR}(x_i^{(k)}, y_j^{(k)} ) \,,
    \label{eqn:obj_hard}
    \end{equation}
    where $\text{XNOR}(\cdot,\cdot)$ returns, element-wisely, whether the $k$-th element of $\mathbf{x}_i$ and $\mathbf{y}_j$ are the same. Alternatively, we replace the $\text{mean}$ aggregator by $\max$, to say it is good enough for $\mathbf{y}_j$ to be close to any $\mathbf{x}_i$.
    \item \textbf{Soft distance}:
    We consider instead the soft grooving patterns $\tilde{\mathbf{x}}_i$ and $\tilde{\mathbf{y}}_j$, and compute the distance between them as
    %\begin{equation}
    $\text{mean}_{i=(1,\dots,N)}  \frac{1}{M} \sum_{j=1}^{M}   \|\tilde{\mathbf{x}}_i - \tilde{\mathbf{y}}_j \|_2^2 \,.$
    %$\label{eqn:obj_hard}
    %\end{equation}
    We can similarly replace mean by the $\min$ function.
\end{itemize}
Table \ref{tab:obj_result} shows that, consistently across different metrics, groove-aware models outperform the no-grooving model. Moreover, the scores of the groove-aware models are closer to the high end than to the low end. It is also important to note that, there is still a moderate gap between the best model's composition and the real data, which has to be further addressed in the future work.

%\subsection{Objective Study}
%We employ the metrics presented in Section \ref{sec:obj_evaluation} to test the rhythmic consistency performance of our generated continuous bar. 
%We also provide the real data and random permutation data to test on this metric in order to set the higher and lower bound of the metric. 
%Table \ref{tab:obj_result} shows that our method all get a better result on the accuracy metric. Especially the multi-resolution hard grooving setting get 7 more percent on the mean hard accuracy metric and get 5.6 more percent on min hard accuracy metric. Multi hard resolution method performs comparably to the real data.

% \begin{table}[t]
% \centering
% %\setlength{\tabcolsep}{5pt} % Default value: 6pt
% \begin{tabular}{l|c}
% \toprule
% Pitch & Error rate \\ 
% \midrule
% 47   & 23.90\%  \\
% 45  & 22.80\%\\
% 44   & 18.57\%\\
% 42  & 17.67\% \\
% 49   & 16.38\%\\
% \bottomrule
% \end{tabular}
% \caption{Result of optional question from high level guitar proficiency participants in first user study.}
% \label{tab:pitch_error_rate}
% \end{table}

% \begin{figure}[t]
% \centering
% \includegraphics[width=\linewidth]{figs/user_study2.png}
% \caption{Result of optional question from high level guitar proficiency participants in first user study.}
% \label{fig:user_study2}
% \end{figure}

%\subsection{User Study 1}
%\label{sec:user_study}

Figure \ref{fig:user_study1} shows the result of the subjective evaluation, where we present the audio rendition (using a guitar synthesizer) of the aforementioned 16-bar continuations to human listeners, and ask them to choose the one they like the most, among those generated by the `no-grooving,' `soft-grooving,' and `hard-grooving' models. We divide the response from 57 participants by their self-report proficiency level in guitar. Figure \ref{fig:user_study1} shows that professionals are aware of the difference between groove-aware and no-grooving models. According to their optional verbal response, groove-aware models continue the prompts better, and generate more pleasant melody lines.

\begin{table}
  \begin{center}
    \begin{tabular}{l|ccc}
    \toprule
       & \textbf{Real} & \textbf{No grooving} & \textbf{ Hard grooving }  \\
       \midrule
       MOS & 3.48$\pm$1.16 & 2.80$\pm$1.03 & 3.43$\pm$1.12 \\
      \bottomrule
    \end{tabular}
    \caption{Result of the second user study (in mean opinion score, from 1 to 5) comparing audio renditions of real tabs and machine-composed tabs by two variants of our model.}
    \label{tab:exp3}
  \end{center}
\end{table}

\subsection{Experiment 3: On Comparison with Real Tabs}
\label{sec:exp3}

%\subsection{User Study 2}
%\label{sec:user_study_2}

Finally, our last RQ involves another user study where we ask participants to rate, on a Likert five-point scale %(1--5; the higher the better), 
how they like the audio rendition of the continuations, this time including the result of real continuations. For groove-aware models, we consider hard-grooving only, for its simplicity and also for reducing the load on the subjects. 
Much to our surprise, the average result from 23 participants (see Table \ref{tab:exp3}) suggests that hard-grooving compositions are actually on par with real compositions.
%The reason we conduct the second user study is we want to know how much closer are our results generated by adding a single token information to the input than the real data. Thus, in the second user study, each study consists of two audio set, each set include 3 audio which are two proposed method and the real data. Beside this, we also ask user to make a rating of each audio in a mean opinion score(MOS) scale. 
%To simplify the study, we also shorten the audio length to 16-bars(i.e., 4 prompt bars and 12 continuous bars). We collect result from 23 participants and show on Figure \ref{} that our method with simply add naive grooving token (i.e., hard grooving setting) are better than the no grooving. Surprisingly, our generated result get the same average value as real data get. 
We believe this result has to be taken with a grain of salt, as it concerns with only fairly short pieces (i.e., 16 bars) that do not contain performance-level variations. Yet, it provides evidence showing the promise of deep learning for tab composition.

% When the following words appear in the conference publication titles, please abbreviate them: Proceedings $\rightarrow$ Proc.; Record $\rightarrow$ Rec.; Symposium $\rightarrow$ Symp.; Technical Digest $\rightarrow$ Tech. Dig.; Technical Paper $\rightarrow$ Tech. Paper; First $\rightarrow$ 1st; Second $\rightarrow$ 2nd; Third $\rightarrow$ 3rd; Fourth/nth $\rightarrow$ 4th/nth.

\section{Conclusion}
\label{sec:conclusion}
In this paper, we have presented a series of evaluations supporting the effectiveness of a modern neural sequence model, called Transformer-XL, for automatic composition of fingerstyle guitar tabs. The model still has troubles in ensuring the note-string association and the rhythmic coherence of the generated tabs.
%, and in realizing a natural progression of the tabs' groove. 
How well the model generates tabs of plausible long-term structure is not yet studied.
%The temporal resolution of the note onsets is only 16th note, 
And, much of the expression in guitar music is left unaddressed. Much work are yet to be done to possibly redesign the network architecture and the tab representation. Yet, we hope this work shows promises that inspire more research on this intriguing area of research.

\bibliography{reference}

\end{document}